\documentclass[sigconf]{acmart}

\copyrightyear{2017} 
\acmYear{2017} 
\setcopyright{acmlicensed}
\acmConference{WebSci '17}{June 25-28, 2017}{Troy, NY, USA}\acmPrice{15.00}\acmDOI{10.1145/3091478.3091488}
\acmISBN{978-1-4503-4896-6/17/06}

\usepackage[english]{babel}

\usepackage[T2A,T1]{fontenc}

\usepackage{booktabs}

\usepackage{tabularx}
\usepackage{ltablex} 
\usepackage{longtable}

\usepackage{supertabular}

\usepackage{placeins}
\usepackage{pgffor}

\usepackage{subfig}
\usepackage{color}
\usepackage{comment}
\usepackage{enumitem}

\usepackage{morefloats} %in order to use more figures
\usepackage{chngcntr} %for figures in Appendix

\usepackage[utf8x]{inputenc}
\usepackage{amsmath}
\usepackage{url}
\usepackage{footnote}

\usepackage[autostyle]{csquotes}  %for block citations
\usepackage{booktabs} %for tables , especially option toprule

\usepackage{hyperref}

\newcommand{\footlabel}[2]{%
    \addtocounter{footnote}{1}%
    \footnotetext[\thefootnote]{%
        \addtocounter{footnote}{-1}%
        \refstepcounter{footnote}\label{#1}%
        #2%
    }%
    $^{\ref{#1}}$%
}

\newcommand{\footref}[1]{%
    $^{\ref{#1}}$%
}

\usepackage{microtype}

\begin{document}

\title{"(Weitergeleitet von \underline{Journalistin})": The Gendered Presentation of Professions on Wikipedia}

\author{Olga Zagovora}
\affiliation{%
  \institution{GESIS - Leibniz Institute for the Social Sciences}
  \city{Cologne}
\country{Germany} }
\email{olga.zagovora@gesis.org}

\author{Fabian Flöck}
\affiliation{%
  \institution{GESIS - Leibniz Institute for the Social Sciences}
  \city{Cologne}
\country{Germany} }
\email{fabian.floeck@gesis.org}

\author{Claudia Wagner}
\affiliation{%
  \institution{GESIS - Leibniz Institute for the Social Sciences and U. of Koblenz-Landau}
  \city{Cologne/Koblenz}
\country{Germany} }
\email{claudia.wagner@gesis.org}

\begin{abstract}
Previous research has shown the existence of gender biases in the depiction of professions and occupations in search engine results. 
Such an unbalanced presentation might just as likely occur on Wikipedia, one of the most popular knowledge resources on the Web, since the encyclopedia has already been found to exhibit such tendencies in past studies.  
Under this premise, our work assesses gender bias with respect to the content of German Wikipedia articles about professions and occupations along three dimensions: used male vs. female titles (and redirects), included images of persons, and names of professionals mentioned in the articles. 
We further use German labor market data to assess the potential misrepresentation of a gender for each specific profession.  
Our findings in fact provide evidence for systematic over-representation of men on all three dimensions. For instance, for professional fields dominated by females, the respective articles  on average still feature almost two times more images of men; and in the mean, 83\% of the mentioned names of professionals were male and only 17\% female.

\end{abstract}


\maketitle

%\category{J.4}{SOCIAL AND BEHAVIORAL SCIENCES}{}

%\terms{computational social science, cultural studies, cross-lingual study, Wikipedia}

\section{Introduction}
\label{sec:intro}

%Wikipedia is an online encyclopedia that is ranked among the six most popular websites on the Internet in August 2016 \cite{_alexa_wikipedia}. According to WikiStats project, Wikipedia has 41.2 million articles \cite{_wikipedia_stats} and had more than 15 billion page views in August 2016 \cite{_wikipedia_views}. Thus, 

%%%%%%%%% observation of problem

An experiment by Chambers~\cite{chambers_stereotypic_1983} conducted in 1983 brought forth an interesting \textbf{observation}: Given the request to draw a picture of a "scientist", only 28 (0.06\%) of 4807 children depicted a female scientist. However, at the time of the study female researchers already made up 14\% of STEM professionals in the United States~\cite{%hill_why_2010,
census_1980}. 
More recently, a study~\cite{kay_unequal_2015} unveiled that a Google search for profession names returns more images of men in the top results than one would expect from the actual gender distribution in the queried profession, i.e., it occurred even if the field was not dominated by men.% For example, in a profession with 50\% women, one would expect expect about 45\% of the images depicting women on average. %These results leading the authors to conclude that not necessarily the search algorithm, but the decision by webmasters how to present many professions on Websites is heavily skewed towards featuring males over females.  

%%%%%% Why is the problem relevant? why professions and gender?

These observations highlight an important \textbf{issue}: Professions and occupations\footnote{In the remainder of this paper, we will refer to both solely as "professions", unless specifically distinguished.} are a central part of societal life; and the mental pictures we hold of these professions shape our opinions about and our attitudes towards them. While this holds true for several perceived features of a person in a profession, like their race or political convictions, gender is one of the most defining characteristic of a stereotype~\cite{greenwald_implicit_1995}, and also one that has sparked great disputes.
%And some professions have  traditionally been associated with male or female domination of the field~\cite{white_occupational_1989}. 

%%%%%%%%%%% effects

%The information people get from online media may affect their interpretations and understanding of the surrounding world. 

\begin{figure}[t!]
    \centering
    \includegraphics[width=\linewidth, trim=0 0 0 0, clip=true]{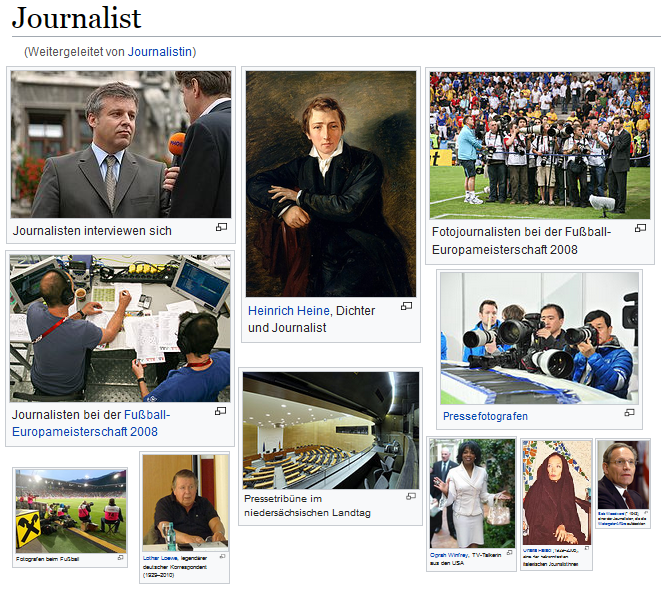}
    \caption[Collage of images]{
    \textbf{Screenshots of all images present in the German Wikipedia article ``\href{https://de.wikipedia.org/wiki/Journalistin}{Journalist}''} as of Feb. 2017.
    }
    \label{fig:collage_wiki}
    \vspace{-5mm}
\end{figure}

Information  that (mis)represents a given state of affairs %(be it webmasters preferring pictures of male doctors or an algorithm ranking them higher)
may endorse existing or create new stereotypes in the reader or viewer~\cite{arendt_effects_2015} which potentially impact their behavior.
Cultivation theory has long suggested that 
 prolonged exposure to biased information may shift individuals’ perceptions over time~\cite{browne_graves_television_1999} and even different gender ratios in image search results affect participants’s perception of gender proportions in professions slightly but notably~\cite{kay_unequal_2015}. 
In respect to professions and gender, a potential \textbf{effect} is that in particular young people can be influenced in their career decisions, after being exposed to such information -- conceivably avoiding an ostensibly (fe)male-dominated trade, even if they have a natural predisposition for it~\cite{correll_gender_2001}.

\textbf{Sources} for such distorted views of a profession can be peers or the mass media - % - but today more than ever content from the World Wide Web.
%Various processes can shape the creation of a certain (gender-)biased presentation of a profession: a deliberate or unconscious bias of the creator(s), a biased filter in the presentation (e.g., a specific ranking algorithm), or simply an out-dated representation of a since-changed ground truth.
%A more modern influence capable of shaping such perceptions is information retrieved from the World Wide Web, e.g., the ranking of results when searching for professions in services like Google
yet today, the Web has arguably become one of the main reference points for personal research, including when acquiring general knowledge about professions. And very frequently, Web surfers end up informing themselves from related \textbf{articles on Wikipedia}, the world's largest and most popular online encyclopedia. 
In 2016 for instance, articles in the category "Occupations" on the English Wikipedia received 368,512,005 unique hits together (of 93 billion in total) and the corresponding category "Beruf" on the German Wikipedia accumulated 30,669,129 unique hits (of 12 billion total)\footnote{According to the  \href{https://tools.wmflabs.org/massviews/}{Wikimedia pageviews API} as of 28.02.17}. %X\% of the German population are using Wikipedia at least once a week, X\% at least "from time to time". In the U.S., these numbers are at x\% and x\% currently.
%37.6\% of search traffic ends up in  
%
%
%% 37.7% of Traffic From Search (Germany) http://www.alexa.com/topsites/countries/DE
%% 37.6% of Traffic From Search (US) http://www.alexa.com/topsites/countries/US
%
%
% 368,512,005 views for last year Cat.Occupations + depth lev.2 %
% 30,669,129 total views for last year Cat. Beruf + depth lev.2 %
%source: http://tools.wmflabs.org/glamtools/treeviews/
%
%Category:Occupations 1/1/2016 - 12/31/2016 
%Totals 	149 pages 	9.781.660 	26.726 / day
%https://tools.wmflabs.org/massviews/?platform=all-access&agent=user&source=category&target=https%3A%2F%2Fen.wikipedia.org%2Fwiki%2FCategory%3AOccupations&range=latest-20&subjectpage=0&sort=views&direction=1&view=list
%Kategorie:Beruf 1/1/2016 - 12/31/2016
%Totals 	64 pages 	490.054 	1.339 / day
%https://tools.wmflabs.org/massviews/?platform=all-access&agent=user&source=category&target=https%3A%2F%2Fde.wikipedia.org%2Fwiki%2FKategorie%3ABeruf&range=last-year&subjectpage=0&sort=views&direction=1&view=list
%
%general pageviews https://tools.wmflabs.org/siteviews/?platform=all-access&source=pageviews&agent=user&range=last-year&sites=de.wikipedia.org|en.wikipedia.org
%
%%%%% or data from here (survey US also but not Germany)
% United States: Between 2.85 and 5.51 pages/week per Internet user (based on actual pageview data) 
%source: http://lesswrong.com/lw/odb/wikipedia_usage_survey_results/#cross-country-comparison-in-perspective
%
Apparently, \textit{a sizable proportion of Internet users is hence  exposed to, and probably influenced by, the presentation of facts about professions on Wikipedia, and the way gender is portrayed in those professionals fields.} 

At the same time, the encyclopedia has already been found to exhibit considerable gender biases in past studies~\cite{collier_conflict_2012,hill_wikipedia_2013,wagner_its_2015} and a similar distortion in its profession articles is entirely possible. 
%in biography articles \cite{wagner_its_2015} as well as a gender gap in the editor community \cite{hill_wikipedia_2013,collier_conflict_2012}.
%
The article ''Journalist`` on the German Wikipedia (Figure \ref{fig:collage_wiki}) is a showcase of how visible gender disparities can become on Wikipedia.
%Gender inequalities are very visible on Wikipedia
%To illustrate the overall issue of gender bias, let us look at some Wikipedia data. As a data point, consider the article “Journalist” on the German Wikipedia (Figure \ref{fig:collage_wiki}). 
Throughout the article, there are many images depicting men and only two images depicting women. Further, considerably fewer female than male journalists are mentioned in the text by name. Lastly, the common female form "Journalistin" doesn't even  exist as an article, instead redirecting to the male form "Journalist".
%A natural implicit or explicit conclusion could be that most journalists are male individuals. This example illustrates that gender bias in Wikipedia may be an important problem and this work sets out to develop a computational approach to quantify such bias and make it transparent.

\textbf{The objectives of our study}, are  hence to reveal (i) how (im)-balanced the gender presentation is on the profession pages of Wikipedia and (ii) test if any imbalance can be explained by underlying labor market data or other facts which could serve to rebut an inherent gender bias of Wikipedia.
We focus on the German Wikipedia as %labor market data is relatively straightforward to obtain, 
it is less studied than its English counterpart and because the language features a large set of female vs. male profession descriptors for analysis, as gender-specific versions exist for almost all profession names.

As illustrated through the "Journalist" example, we focus on three dimensions of profession articles: we (i) study redirections\footnote{A redirection automatically forwards from a requested lemma to the redirection target, another lemma/article. At the target page, a tiny notification "redirected from X" (EN) or "weitergeleitet von X" (DE) appears, cf. top of Fig.\ref{fig:collage_wiki}.} between neutral, male and female profession descriptors, and we analyze (ii) the ratios of male and female images as well as (iii) the ratios of mentioned male and female names of professionals.
%\noident The \textbf{objective} of this work is to identify and assess gender bias related to professions in Wikipedia articles. More precisely,  explore gender bias on 3 dimensions: gender inclusiveness in the use of titles and corresponding 
%%%%%%% contributions
Our work shows that for most professions for which a male and a female title exists in the German language, only the male title has a corresponding article.
In the articles, almost four times more images depicting men than women were encountered. Articles about professions also tend to mention more men. 
In fact, 75\% of all articles have proportions of mentioned men from 0.8 to 1.0.
Moreover, 83\% men and only 17\% women were mentioned on average in the articles. To explain the male bias in the German Wikipedia, we compared it with the Google bias and offline imbalances in the labor market, indicating that  Wikipedia not only inherits gender inequalities from other sources but also aggravates skewed gender ratios in many cases.

\section{Data collection}\label{sec:datasets}

\noindent In order to conduct our analyses, several datasets had to be collected and prepared.

\textbf{Seed dataset of profession names.} An official list of professions from the German Federal Employment Agency (“Bundesagentur für Arbeit”)\footnote{Retrieved from \url{http://berufenet.arbeitsagentur.de/berufe/berufe-beschreibungen.html} on 15.06.15%, alternative link \url{https://berufenet.arbeitsagentur.de/berufenet/faces/index?path=null/sucheAZ&let=A}%, file can be accessed from \url{https://www.dropbox.com/s/d8w20gbpgfzm7l9/Berufsbezeichnungen.txt?dl=0}
} was retrieved in order to create a comprehensive seed list of corresponding male-female pairs of profession names.

We parsed the profession descriptors using the grammatical rules of the German language including corresponding suffixes in order to create appropriate male and female forms of job titles. 
4274 pairs were generated, such as "Lehrer"-"Lehrerin", "Krankenpfleger"-"Krankenschwester“, "Entbindungspfleger"-"Hebamme" etc.  Further, we extracted 183 gender-neutral profession labels from the initial professions list, e.g., "PR-Fachkraft", "Fotomodell", "Aufsichtsperson". The complete lists were  manually cross-validated by two of the authors. 

\textbf{Wikipedia profession articles dataset.} 
To map the seed list of  profession names to  Wikipedia  articles 
%lemmas (encyclopedic entries with a URL and title)
we extracted all articles 
%(including such that only contain redirects) 
belonging to the German Wikipedia categories “Profession” (DE\footnote{Henceforth  ``DE'' is used as an abbreviation for "German".}: “Beruf”), “(Public) position” (DE: “Amt”), “Person by occupation” (DE: “Person nach Tätigkeit”), and any of their subcategories down to the 5th depth level.\footlabel{note1}{Accessed on 07.02.16 from \url{https://de.wikipedia.org/w/api.php}} This was done as (i) profession labels can be homonyms and hence match a seed list entry to a non-profession article and (ii) to find a comprehensive, yet still clearly profession-related set of articles to compute a more relaxed matching method with.
We then constructed the intersection of the set of profession names from the seed list with the profession article title set, by applying a relaxed string-matching: the Levenshtein distance and ratio (the match proportion of two words) were  calculated  between  each  profession  name  and  each  article  title.  
If the Levenshtein distance was at most 2 or the Levenshtein ratio was at least 0.8, the corresponding pairs of words were matched. Hence, we were able to find profession articles with titles written in a slightly different  manner. %than  in  the  original  list  of  professions.
In  order  to  avoid  inappropriate  matches (e.g., profession "B\"{a}cker" and name "S\"{a}cker"), all matched pairs were manually validated. %Then the new name (article title) was manually assigned to male, female or neutral group of profession names. 
We proceeded to add 22 professions from the seed list that we identified by hand to have a matching article in Wikipedia, but lacking the appropriate category label.

For each matched article title, additional information regarding its redirection was stored, i.e., whether the original lemma  redirected to another article - and to which one.  

%We  then validated  if  the  matched articles  are  indeed about  professions.  For  this  purpose,  all  categories  an  article belongs  to  were  examined.  If  the  article  belongs  to  one  of  the  following    Articles not contained in those categories were discarded. %Consequently,  we  restricted  all  matched  articles  only  to  those  which  are  about professions. 
 
%To extend the set of matches, the following semi-automatic method was applied.  
%The  Levenshtein  distance  measures  the  minimum  number  of  single-character  edits required  to  change  one  word  into  the  other  \cite{levenshtein_binary_1966}.  The  Levenshtein  ratio  between  words  a  and  b  is defined as follows: 
% $$ Lev$$
% where  LevDistance(a, b)  is the Levenshtein distance between words a and b,  $len(a)$ and $len(b)$ define the length of words a and b. Thus, the Levenshtein ratio reflects the match proportion of two words.

%Since the matched articles were taken from profession (and occupation) categories, there is no need to further validate them to be about professions. 
 
As a result, a high-quality list of unique, non-redirect Wikipedia articles (885 entries) about professions was collected.\footnote{One reason for the relatively small overlap is that the seed list is  exhaustive and contains many specialized professions that seem not to be "notable" enough for the DE Wikipedia.}
An additional 820  lemmas that matched with the seed list were redirects.

\textbf{Gender-specific person names.} 
To identify if and how many female or male persons are mentioned in a profession article,
%The dataset of people, which are mentioned in the Wikipedia articles about professions, 
we applied a two-pronged approach. % through two separate methods. 

First, all internal links %(including the linked text)
in the profession articles were collected which point to articles about persons either in the Wikipedia-categories “Woman” (DE: "Frau") or “Man” (DE: "Mann") using the MediaWiki API\footref{note1} 
%(Link Dataset)
- these were each counted as one mention of a person with the Category-derived gender.
%Thus, names of people within a Wikipedia article were collected (LinkDataset).

Second, the Named Entity Recognition method proposed by Al-Rafou et al.~\cite{al-rfou_polyglot-ner:_2015} was applied to the complete article text content, yielding all Entity Names of class “Person” that were identified.
Then, for each found person mention, the gender was identified according to the first name. To make gender identification more accurate, several vocabularies \cite{karimi_inferring_2016,michael_2007}  were used. %(Polyglot Dataset).
The results obtained by these two methods were combined for each article, producing an overall dataset of 5085 identified person labels (4272 men and 813 women). %We manually assessed and determined such 
For 
cases where the two datasets clashed in the determination of the gender (2.77\%), we were able to identify the correct genders automatically since persons from the first dataset are readily gendered. 
The end-result allowed us to reliably assess the number of female or male individuals mentioned.
%By combining the datasets, we can also estimate the error rate (2.77\%) of the gender determination of those persons from the PolyglotDataset who are also in the LinkDataset. This was possible since persons from the LinkDataset are readily gendered.

\textbf{Images of persons.} The MediaWiki API\footref{note1} was used  to retrieve all images contained in the articles. Only images wider than 100 pixels were stored, assuming that small images are either icons or too small to recognize the gender. 
A manual corroboration of this assumption with about 80 pictures  yielded zero false positives for this filter.
Files with the following formats were also excluded:
%Files of “svg”, “ogg” and “ogv” formats were excluded, since
“svg”-files are vector images used solely for schemas and icons, and “ogg”, “ogv” are video and multimedia text formats. Thus, 906 images  were collected from 345 profession articles. The remaining articles did not contain (suitable) images.

\textbf{Labor market statistics.} Gender-specific employment statistics %\footnote{German: “Beschäftigungsstatistik”}
were obtained from the “Statistics of the Federal Employment Agency” (DE: “Statistik der Bundesagentur für Arbeit”). The statistics consist of absolute numbers of men and women involved in profession subgroups as of June 30, 2015. Some examples of profession subgroups are: “8445; (Fremd-) Sprachenlehrer/innen” and “8442; Berufe in der Religionspädagogik”.
Each profession in our seed dataset is annotated with a specific classification number by the Employment Agency, enabling the assignation to its respective profession group according to an accompanying profession directory (DE: “Klassifikation der Berufe 2010 - alphabetisches Verzeichnis der Berufsbenennungen”)
%, i.e., by using the profession encoding number 
E.g., “8445x” is the encoding of all professions in subgroup “(Fremd-) Sprachenlehrer/innen”. %Then, for each profession, the percentage of women in the profession was estimated.
In this way we could match 871 of the 885 article-mapped professions to their respective labor market statistics including gender distributions.

\textbf{Google hits.} By using each of the female and male profession labels as a search query term, we collected the amount of Google search results (hits) generated through the Google Web Search API\footnote{Accessed on 09.02.16 from \url{http://ajax.googleapis.com/ajax/services/search/web?v=1.0&hl=de&btnG=Google+Search&q=}; the API results are not subject to custom personalization.  %project immigrated to \url{https://developers.google.com/custom-search/}
} to gauge the number of Web resources featuring these descriptions. 
%
%Results were retrieved from the API because numbers of Google hits from www.google.de are not reliable. That is, the number of hits can vary, since it depends on custom settings such as used browser, location of user, and Google profile of user.
The search scope was restricted to the German language using the corresponding parameter in the API query, as some search terms could also be valid English words, for example.

\section{Research Method}\label{sec:methods}
To assess whether and to what extent the German Wikipedia exhibits a gender imbalance with respect to professions, the collected articles have been analyzed along three dimensions: (i) gender inclusiveness in job titles and a corresponding redirection analysis, (ii) analysis of male-female image proportions,  and (iii) the balance of male-female mentions. We present these dimensions in the following subsections. 
 
\subsection{Redirection analysis}\label{redirection_method}

The following analysis was inspired by the  scenario of a user wanting to inform herself about a specific profession  and entering the respective (fe)male form in Wikipedia's search bar or directly navigating to {\sf de.wikipedia.org/wiki/<profession\_title>}. %\footnote{Another scenario: The search for the (fe)male form on Google, where Wikipedia results often occupy top result ranks - and only the opposite gender form might appear.} 
We were interested in how many cases such a search will lead to either no results or a redirect to the opposite gender form or a neutral page -- with redirects happening immediately after visiting a target article address, being indicated with an unobtrusive disclaimer, and hence possibly going unnoticed.  

For each of the female-male pairs in our profession list (and the neutral labels), we applied these rules to sort each profession into distinct classes, indicating a potential skewness, or bias\footnote{Note that we do not associate ``bias'' here with any deliberate attempt to systematically discriminate, cf. discussion.}, towards men or women: 
\vspace{-1mm}
%
%The study reveals how many and which professions are represented by both Wikipedia articles (with male and female job titles) and how many professions are represented at least by one. 
%
%If a profession has two articles with the male and corresponding female title (as indicated by our datasets) we assume that such a presentation of the profession is not skewed towards men or women, i.e., there is no bias (note that we do not associate ``bias'' here with any deliberate attempt to systematically discriminate, cf. discussion). Therefore, that profession will be associated with a “neutral” representation. If a profession  only has a Wikipedia article with a male title, the profession will be associated with a “male bias”; if  
%the other way around, i.e., 
%it is only featured in  an  article with the female title, the profession will be associated with a “female bias”. 
%
%If a \textit{redirect} takes place then the profession will be associated with (i) the “male bias” group in case of redirect from female to male job title and (ii) with the “female bias” group if the redirect occurs in the opposite way. 
%The third case is when a profession name redirects to an article with neutral name of a profession or is only represented by a neutral title. We  assign such professions to the “neutral” group. %Professions which are only represented via articles with neutral profession names were also associated with the “neutral” group. 
\begin{enumerate}%[itemsep=-1mm]
\item If articles/lemmas do not exist at all on Wikipedia for a profession in our seed list -- \textit{no evidence};
\item If (non-redirect) profession articles with both gender titles exist (i.e., male and female titles) -- \textit{neutral};   
\item If a profession article with a male / female title exists without the other gender form existing as an article -- \textit{male bias} or \textit{female bias};
\item If only an article with a neutral title exists -- \textit{neutral}.
%\item If a profession article with only a female title exists -- female bias;
\item If a profession lemma exists, but redirects to the other gender -- bias in favor of the redirection target (i.e., \textit{male bias} or \textit{female bias});
\item If a profession lemma with a male or female title redirects to the neutral form or field name -- \textit{neutral};
%\item If a profession article with a female title automatically redirects to the neutral form or field name -- neutral;
\item If a profession lemma exists, but redirects to some article whose title is not in our seed dataset -- \textit{other}. 
\end{enumerate}

\noindent After assigning professions to these bias groups%, one can assess their size. In other words
, we can assess whether professions are more likely to be presented via male, female or neutral profession names. 
%For instance, if one observes more professions in the male bias group, it means that there are more professions which are represented only via articles with a male title in Wikipedia.
However, the non-existence of one gender form can have a host of reasons, based, e.g., on the fact that virtually no (wo)men are employed in a specific profession. Also, we were interested how idiosyncratic the profession representation of Wikipedia is in comparison with the Web in general. 
For the latter, we compared the amount of results found through the Google search engine for each profession label to Wikipedia. For the former, we made use of the German labor market statistics.

%ofIn this next we examine if predominance of one bias group can be explained by popularity of profession names on the Web or by German labor market statistics. That is, Wikipedia could reflect and inherit gender bias from other media or gender inequality of labor market.

\textbf{Google hits.} 
%Initially, we investigate whether phenomena observed on Wikipedia are specific to Wikipedia or if they can be explained by popularity of profession names on the Web. 
As a starting hypothesis, we presume that profession titles that appear on Wikipedia are more popular on the Web, meaning that there are more sources on the Web about them than about the corresponding profession titles of the opposite gender. %, leading to a slanted representation on-Wiki. 
Consequently, one should observe more search hits for profession titles represented on Wikipedia than for those that are not.
As a proxy, we look into the number of hits returned by the Google search engine. %To this end, the Google Web Search API was used, which, besides ranked results also returns the number of Google search results for a specific query. 
For each profession, the number of hits was stored for female and male job titles.
In order to assess if the difference between Google hits for female and male titles is significant, a two-sided %Wilcoxon-
Mann-Whitney rank-sum test was utilized.% between hits for female and male job titles. 

Next, we examined whether one can describe the relationship between the number of Google hits and the redirection bias of a profession. First, for each profession, the normalized difference between hits for male and corresponding female job titles was estimated %(adjusting for overall popularity of a  professions) 
using the following formula:
\begin{equation} Normalized\_di\textit{fference}_i=\frac{Hits_{male_i}-Hits_{female_i}}{Hits_{male_i}+Hits_{female_i}}\end{equation}
where $Hits_{male_i}$ is the number of hits returned for male title of profession $i$, $Hits_{female_i}$ is the number of hits returned for the corresponding female title of profession $i$. 
A positive difference indicates that more search results have been returned for the male profession title than for the corresponding female title, while a negative difference indicates the contrary. The difference approaches zero if both gendered profession titles returned the same amount of search results.

%Then, we checked the possibility of predicting the redirection bias using Google hits for profession names. As described earlier, the professions were classified into three redirection bias groups: male, female, and neutral. In the next step, we fit models which predict the redirection bias group of profession.
%
Then, %Two
two logistic regression models were fitted 
in order to predict the redirection bias using Google hits for profession names. 
The first model uses  the state of being in the male bias group as a dependent variable; the second model deals with female bias instead. %In other words, the models will predict whether a profession has male or female bias. 
The regression functions for both models are given by:
\begin{equation}
p_i= \frac{1}{1+e^{-(\beta_0+\beta_1 x_{1,i} + \beta_2 x_{2,i})} } 
\end{equation}
where $x_{1,i}$ is the normalized difference of Google hits of profession $i$, and $x_{2,i}$ is the Google hits for male title of profession $i$. %In other words, independent variables in both models are the following: the normalized difference of Google hits; Google hits for the male job title.

\textbf{Labor Market.} 
%This study examines whether phenomena observed on Wikipedia can be explained by labor market statistics. One could hypothesize that professions with more women might only have a female article, whereas professions with more men might only have a male article. 
%
For the next analysis, %The statistics consist of numbers of people involved in profession subcategories according to a classification of professions (German: “Klassifikation der Berufe 2010”). Each profession was associated with an appropriate profession subcategory according to the aforementioned classification. Thus, each
each profession was coupled with the corresponding number of employed people per gender as per the labor market statistics collected. % Since some of the professions are too ambiguous (e.g., occupation “Leiter”), there is no labor market statistics for them. Statistics from the German labor market statistics were associated with 859 out of 873 professions existing on Wikipedia.
For each profession, the percentage of women involved was obtained. Then, the dependence between the percentage of women involved in the profession and the redirection bias group of the profession was tested. The null hypothesis is that two sets of measurements are drawn from the same distribution, i.e., the percentage of women involved in profession of the first bias group is drawn from the same distribution as the percentage % of women involved in the profession of 
from the second bias group. The alternative hypothesis is that values in one sample are more likely to be larger than the values in the other sample. 

The Wilcoxon rank-sum test was used for each pair of the bias groups (male-female bias, male-neutral bias, neutral-female bias) in order to find profession groups which show significant differences in the percentage of employed women. Bonferroni correction \cite{dunn_estimation_1959} was used in order to control for the family-wise error rate.
To describe the relations between the bias groups using the percentage of women involved in professions, logistic regression models were fitted. Analogously to the analysis of Google results, we predict whether a profession is in the female\textbackslash neutral\textbackslash male bias group. The explanatory variable is the percentage of women involved in a profession. Since not all bias groups show significant differences between distributions of values, we fit our models only between those which do.

\subsection{Images analysis} 

To identify the gender of people on an extracted picture, a CrowdFlower\footnote{\url{https://www.crowdflower.com/}} task was set up. %Interfaces designed for the CrowdFlower task are presented in Figure 2. 
First, CrowdFlower workers were to identify whether the image shows one or several persons, or none. %Depending on the number of depicted people, the  workers received slightly different questions.  % by choosing one of the following answers: “No Persons”, “One Person”, “Several Persons, *no single* person’s depiction is dominant”, “Several Persons, *but one* person’s depiction is dominant”. 
If the image depicted more than one person, workers were asked to identify whether s/he was depicted in a dominant way, as 
%We add this constraint since %images can depict several people and 
we are interested only in the gender of the main person. %Figure 3 represents several examples of Wikipedia images which depict more than one person, but one person’s depiction is dominant.
For every image with one person
or where one person is dominant, workers were asked to identify the gender %(Figure 2a) 
of that individual. %For every image with more than one person where one is dominant, workers had to identify only the gender for that dominant person. %(Figure 2c)
Otherwise, they were asked to identify the gender of the majority of people on the photo. %(Figure 2b)
For non-recognizable gender or an equal number of men and women, corresponding options were offered. %(Figure 2b)

%The images were grouped in a way that: 
Images depicting one (fe)male, one dominant (fe)male, and images with (fe)male majority were assigned the label ``female'' and ``male'', respectively. %; images depicting one female, one dominant female, and images with female majority were assigned to one single image group “female”; 
Images where gender was not recognizable and with equal number of men and women were assigned two separate labels.

Each image was classified by at least three different workers. In order to control the reliability of all responses, the CrowdFlower accuracy %(the percentage of correct answers)
threshold for workers was defined as 70\% in the setting of the  task. %Further, Fleiss' kappa score was utilized to gauge the consistency of answers for the whole project.
%We also investigated how consistent the responses of the workers were for the whole project; . 
We also manually labeled 15\% of the images such that one of 10 images shown to a CrowdFlower contributor would be from the labeled set - this was used as an additional worker accuracy control.
%Since we know the “right answer” for the labeled images, the accuracy of contributors is assessed based on their answers for those images. %In other words, accuracy score is the percentage of correct answers for the labeled images.
%When a contributor starts our CrowdFlower job his or her accuracy score equals 100\%. 
%The accuracy score is recalculated each time answers for a 10 image set are received.
%after receiving answers for 10 images by CrowdFlower system.
%If the answer of the contributor does not match “the right answer”, contributor’s job accuracy goes down.
If the accuracy of a worker fell below the accuracy threshold, the contributor would be removed from the job and  her answers would not be taken into account. %Moreover, in order to start CrowdFlower task, workers should pass the test tasks, where all 10 images are chosen from the labeled set.

Further, for the whole CrowdFlower task the worker agreement per Fleiss' kappa \cite{powers_evaluation_2011} was estimated.
\begin{comment}

using the following formula:
\begin{equation} \kappa=\frac{\overline{P}-\overline{P_e}}{1-\overline{P}}
\end{equation}
where  $\overline{P}$  refers to the mean of $P_i$'s and $P_i$ refers to the extent to which workers agree for the $i$-th image. In other words, $P_i$ defines how many worker--worker pairs are in agreement, relative to the number of all possible worker -- worker pairs.

and is calculated using the following formula: 

\begin{equation}
P_i=\frac{1}{n(n-1)}[\sum_{i=1}^k{(n^2_{ij})} -n ]
\end{equation}
where $n_{ij}$ refers to the number of raters who assigned the $i$-th image to the $j$-th category, $n$ refers to the number of answers per image (i.e., three in our case), $k$ refers to the number of categories.
\end{comment}
This allowed us to measure the degree to which the observed amount of agreement among workers exceeds what would be expected if all workers made their choices completely randomly. %In other words, it can give us a clue on how consistent responses of the workers are. 
General reliability of agreement between CrowdFlower contributors was achieved with a 0.75 Fleiss’ kappa score. %Overall, the Fleiss’ kappa score is considered to be moderate in range 0.41-0.6, substantial in range 0.61-0.8 and almost perfect agreement in range 0.81-1.00. 
Thus, we can assume that within our task relatively high agreement was achieved.

After gathering all answers for all images, the majority answer was used to label the image. 

%separate groups “gender is not recognizable” and “mixed, equal amount of male and female”. 

\textbf{Labor Market.}
In order to test whether images from the profession articles reflect labor market statistics,
% the following analysis of images was performed.
professions were divided into two groups:
 professions with female ($> 50$\% women) and male majority according to the labor market statistics. 
%Thus, professions with more than 50% women were in one group and professions with more than 50% men were in another group. 
Then, the statistical significance of the difference between distributions of image groups were tested using chi-square
 independence tests with Monte Carlo p-value simulations \cite{north_note_2002}.\footnote{The simulations were used since some image categories have small numbers,
 possibly making p-values unreliable.} 
%The null hypothesis is: category of image and gender of majority in profession are independent. 
%The alternative hypothesis is: category of image and gender of majority in profession are not independent.
We also explored how distribution of image categories would look like if one restricts professions 
to those with more than 70\% men or women, respectively. 
%Analogously, chi square test was used in order to test whether the difference is significant.

Next, the strength of relation was examined between the number of images depicting 
a particular gender in the article and the labor market statistics of profession. 
The Spearman’s rank correlation %(with correction for ties \cite{taylor_correcting_1964})
was utilized. 

We also tested whether distributions of image categories are significantly different
 for articles with male, female and neutral titles. 
 The images were grouped according to the gender of article titles and chi-square tests were applied.
 %. Then chi-square independence tests with Monte Carlo p-value simulations were performed. 
%Thus, we test the statistical significance of the difference between distributions of image categories of these article groups. The null hypothesis is: category of image and gender of article title are independent. The alternative hypothesis is: category of image and gender of article title are not independent.

Analogously, we analyzed whether distributions of image categories are significantly different 
for professions which were assigned to different redirection bias groups. 

%If the test showed a statistically significant difference in the image composition between different article/profession groups, the groups which show the significant difference were revealed 
% and post-hoc tests for them were performed to find image category which shows the significant difference. 
 %. For this purpose pair comparisons (i.e., post-hoc tests) of all groups were performed. 
% To control for the family-wise error rate, the two stage p-value correction of Benjamini-Hochberg\cite{macdonald_type_2000} was utilized.

\subsection{Mentioned people analysis}

%The dataset of person names was used for this analysis.
Articles that did not mention any persons per our name dataset were excluded from the analysis, leaving 411 articles. Then, for each article, the proportion of mentioned men was calculated. 

First, the dependency between the proportion of mentioned men and the gender of the article title was tested, by examining if %:
%a)	articles with male title have higher ratio of mentioned men than articles with neutral title,
%b)	articles with male title have higher ratio of mentioned men than articles with female title,
%c)	articles with neutral title have higher ratio of mentioned men than articles with female title.
articles with male title have a higher proportion of mentioned men than articles with female titles. 
To this end, three %Wilcoxon 
rank-sum tests (i.e., one for each pair of article groups) 
with p-value correction \cite{macdonald_type_2000}
were performed. 
%The following data was compared: ratios of mentioned men in articles with female and male title; ratios of mentioned men in articles with female and neutral title; ratios of mentioned men in articles with male and neutral title. The null hypotheses are that two sets of ratios of mentioned men are drawn from the same distribution. The alternative hypotheses are that values in one set are more likely to be larger than the values in the other set. The two stage p-value correction of Benjamini-Hochberg was used in order to control for the family-wise error rate. 

Second, we tested the dependency between the proportion of mentioned men and redirection bias group of profession. Thus, analogously to the previous analysis, the professions were grouped and then %Wilcoxon
rank-sum tests were applied between proportions of mentioned men in each of the groups. %According to the redirection analysis we have three redirection bias groups (male, female and neutral), thus it requires three tests and p-value correction. Analogously, the two stage p-value correction of Benjamini-Hochberg was applied for the family-wise error rate control.

\textbf{Labor Market.}
Third, we examined whether the proportions of mentioned men in the profession articles reflect labor market statistics. The professions were divided in two groups: professions with more than 50\% men and %professions with 
more than 50\% women. If proportions of mentioned men in the profession articles reflect the labor market, one would observe a high proportion of mentioned men in the group of professions with male majority and low proportion of mentioned men in the group of professions with female majority. %We examined whether the difference is significant between the groups of professions using %Wilcoxon
%rank-sum test. 
Next we analyzed the strength of the relation between the percentage %/number
of mentioned men/women in an article and the percentage/number of employed men/women in a profession. %Thus 
To this end,
the Spearman’s rank correlation was %calculated between:
%•	the percentage of mentioned women in profession article and the percentage of employed women in the profession;
%•	the number of mentioned men in article and the  number of employed men in the profession;
%•	the number of mentioned women in article and the number of employed women in the profession;
%•	the number of mentioned men in article and the number of employed women in the profession;
%•	the number of mentioned women in article and the number of employed men in the profession;
%•	the number of mentioned men in article and the number of employed people in the profession;
%•	the number of mentioned women in article and the number of employed people in the profession.
utilized. 

%Wikipedia articles usually reflect not only on current state of profession, but also the historical past of it and famous people involved in different points in time. Therefore, the mentioned people in articles were limited to those who reflect current labor force. The pool of people was restricted to those who were born after 1960. In order to get dates of birth of mentioned people, the SPARQL queries on DBpedia were applied. The predicate “birthDate” values of persons, who were mentioned in articles, were queried. We were able to find DBpedia objects according to their article id in Wikipedia, i.e, the scope of mentioned people was limited to those with a Wikipedia article. For persons who do not have DBpedia entry, the first line of their Wikipedia articles was parsed in order to retrieve birth date.	
\section{Results}\label{sec:results}
\noindent Our results on gender inequality in Wikipedia's profession presentations are outlined for the three distinct dimensions. %using the approach which we described in the previous section.

%================================
\subsection{Redirection analysis}

%In the set of 4457 (4274 male/female and 183 neutral) professions, 991 Wikipedia articles and 820 redirects %(Table 1a) were encountered. 
%Further analysis of the Wikipedia articles revealed that only 869 articles were about professions, among them are the following: 815 profession articles with male title, 18 profession articles with female title and 36 articles with neutral name of profession. %(Table 1b). 
%Overall, 885 articles about professions were found using our matching approach. 
%One can see from Figure 4 that most
Most of the pages (885) about professions we found on Wikipedia have male titles (831), compared to much fewer female titles (25). A few articles about professions have neutral profession names (36). Hence, at first glance, the Wikipedia community is more male profession-oriented.
For corresponding male-female profession title pairs , we encounter eight ``neutral'' professions (i.e., 16 articles) that have corresponding articles for both gender versions. % female and male names of professions. These eight professions according to our classification rules. %(see Subsection \ref{redirection_method}). % were assigned to the neutral group. 
Articles with a neutral profession title were also assigned to the neutral group.

Among the 820 redirects, %(Table 1c) 
%503 redirects from male label of profession, 310 redirects from female label of profession and seven redirects from neutral label of profession. 
5 are from the male to the female label of a profession, 214 redirect from female to male labels, and 3 redirect from neutral to male labels; other redirects are either to broader fields of professions or titles that are synonyms.
For example, going to the Wikipedia article ``\href{https://de.wikipedia.org/wiki/Sekretärin}{Sekretärin}'' (EN: female secretary), will  automatically redirect to the article ``\href{https://de.wikipedia.org/wiki/Sekretär}{Sekretär}''(EN: male secretary), and thus, one never reaches a Wikipedia page ``Sekretärin''. 
%Female professions that redirect to male professions or have only male profession articles were assigned to male bias group. Thus, we identified 807 professions that have only male profession article and five more were found from the set of articles with female title that redirect to articles with corresponding male title. Male professions that redirect to female professions or have only a female profession article are identified as female bias group (six cases). Professions were also assigned to the neutral bias group if the following conditions were encountered simultaneously: profession has only one article (either with male or female title) and corresponding job title of opposite gender redirects but not to articles with male or female job title (11 cases). 

%One can see from Figure 5 and Table 1c a strong imbalance in the number of redirects between groups of female, male and neutral profession names. Redirects from female to male profession articles are more common than from male to female. Pages with neutral profession names do not redirect to articles with female title and there are three cases with redirects to articles with male title.

As a bottom line, the combination of redirects and existing pages of corresponding male-female profession names reveals 812 male bias cases, 6 cases of female bias, and 55 neutral cases. I.e., 812 professions have only an article with male title or a female title lemma redirect to an article with the corresponding male title, 6 professions have only an article with female title or a male title lemma redirect to the corresponding female title, 55 professions have either an article with neutral title or articles with both male and female titles. Thus, we observe evidence for gender disparity among article titles chosen by the Wikipedia editors.

%Next, we wanted to know whether the observed preference of male titles over female titles is a specific phenomenon of Wikipedia. Therefore, we firstly explored how popular male and female professions are on the Web and secondly investigated their popularity among men or women according to the labor market statistics.
\begin{figure}[t!]
    \centering
    \includegraphics[width=\linewidth, trim=20 20 32 27, clip=true]{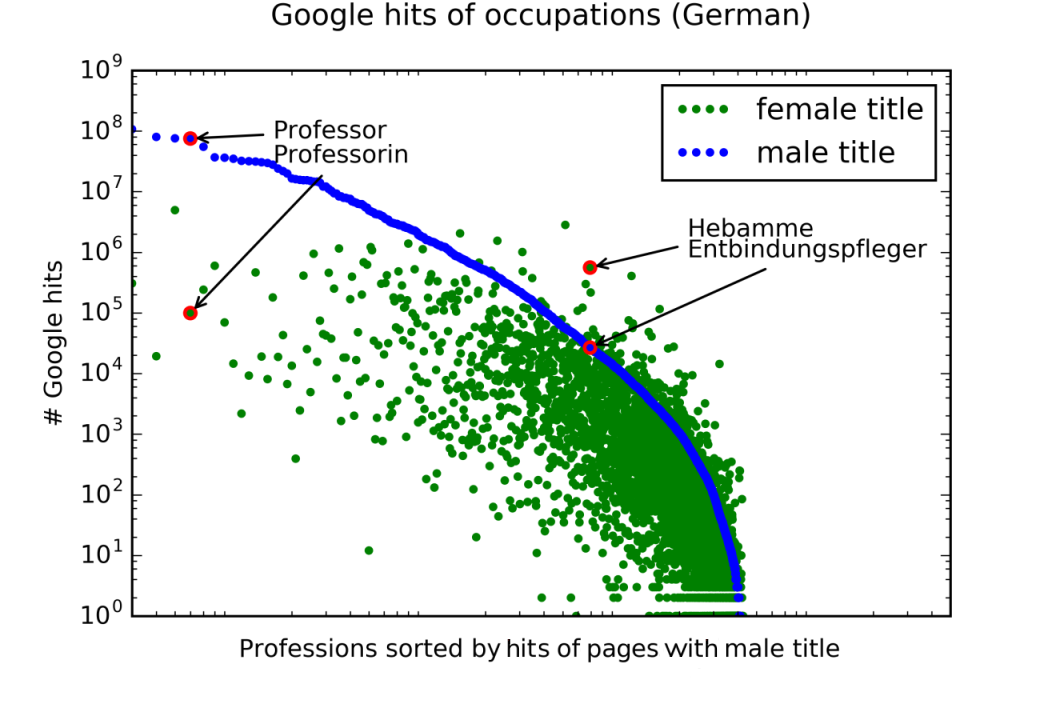}
    \caption[Distribution of Google hits ]{\textbf{Distribution of Google hits} 
    for profession names, sorted according to amount of hits (log) for male profession names. One can see that most female  titles have fewer Google results than corresponding male titles. For instance, ``Professor'' has more Google hits than the female  ``Professorin''. The female profession names are more popular than the male ones for only a few professions, e.g., ``Hebamme” (EN: midwife) vs. “Entbindungspfleger” (EN: male midwife). 
    }
    \label{fig:google-hits-distr}
    \vspace{-4mm}
\end{figure}

\textbf{Google hits.}
%To examine whether the phenomenon observed on Wikipedia can be explained by the popularity of profession names on the Web, we looked into the Google hit statistics (i.e., the number of search results found) for male and female profession titles. 
%We first hypothesize that professions for which a male and female title exist will have a Wikipedia article for the title which is popular on the Web, independent of gender.
%for the male title if this title is more popular on the Web or the female title if this one is more popular
%Our hypothesis is that profession titles with a Wikipedia article are more popular on the Web than corresponding profession titles of opposite gender without a Wikipedia article. For example, one would observe more sources for male profession titles than corresponding female titles for those professions with only a male article on Wikipedia (i.e., male bias group) and similar trend for professions with a female article. %One would also observe more sources for female than male titles for professions with only a female article on Wikipedia (i.e., female bias group). 
%This would mean that choices of Wikipedia editors reflect the popularity of profession names on the Web.
%
%First, we compared the number of Google hits for male and corresponding female names of professions. 
Figure \ref{fig:google-hits-distr} shows the distribution of Google hits for the profession names in our dataset. One can see that male profession names tend to be more popular than female names ($z = 28.6, p^{***}%$p<<0.0001 
$) on Google. 
%Most female profession names have fewer Google hits than the corresponding male profession names,
This indicates that the German speaking Web features more sources for male than female profession names, potentially reflected on Wikipedia.

%According to the results of Wilcoxon rank-sum test between distributions of Google hits of male and female profession names, the difference is significant ($p<<0.0001, z = 28.6$) and Google hits of male profession names are more likely to be greater than respective values of female profession names. 

Next we explored the normalized difference between Google hits of male and female profession titles for 
three groups separately: all professions with male, female, and neutral redirection bias identified previously. 
Our results indicate that a logistic regression model (Table \ref{log_regr_google}) can indeed predict if a profession has a male or female redirection bias, based on  the normalized difference of Google hits.
The coefficients reveal that with a one-unit increase in the normalized difference of Google hits, odds of a profession having a male bias on Wikipedia increases by a factor of 11.48, whereas the probability of having a female bias decreases. In other words, it is more likely that professions have only a Wikipedia article with a female profession title (female bias group) when the female profession name has a greater number of Google hits than the male profession name. %The results of the logistic regression models fitting are shown in Table 2. 
This indicates that the male skew of the German speaking Web is indeed mirrored on Wikipedia.

%\begin{figure}[t!]
%    \centering
%    \includegraphics[width=\linewidth, trim=1 0 2 0, clip=true]{figures/log_regr_google.png}
%    \caption[Logistic regression Google hits]{ Results of the best fitted logistic regression models. Model 1 stands for logit model with binary outcome of profession being in the female bias group or not. The coefficient for “Normalized Google difference” reveals that, we will see 11.48 factor increase in the odds of being in a female bias group for a one-unit increase in “Normalized Google difference” score, since $exp(2.44) = 11.48$. Model 2 stands for logit model with binary outcome of profession being in the male bias group. The coefficient for “Normalized Google difference” reveals that, we will see 0.0026 factor increase (or $1/0.0026=384$ factor decrease) in the odds of being in a male bias group for a one-unit increase in “Normalized Google difference” score since $exp(-5.93) = 0.002654$. For example, both regression models will predict a female bias when the value of normalized Goggle difference equals -1 and male bias when the value equals 1.    }
%    \label{fig:log_regr_google}
%\end{figure}

\begin{table}[t!]
\centering
    \caption[Logistic regression Google hits]{\textbf{Logistic regression for Google hits.} 
    (Results of the best fitted logistic regression models) Model 1 stands for a logit model with binary outcome of a profession being in the female bias group or not. We see an 11.48 factor increase in the odds of being in a female bias group for a one-unit increase in the ``Normalized Google difference'' (NGD) score, since $exp(2.44) = 11.48$. Model 2 is the logit model with the binary outcome of a profession being in the male bias group. The coefficient for NGD reveals a 0.0026 factor increase (or $1/0.0026=384$ factor decrease) in the odds of being in a male bias group for a one-unit increase in the NGD score since $exp(-5.93) = 0.002654$. For example, both regression models will predict a female bias when the value of normalized Goggle difference equals -1 and male bias when the value equals 1.  
    }
\label{log_regr_google}
\resizebox{\columnwidth}{!}{
\begin{tabular}{|c|c|c|c|}
\hline
    & coef.  & p & 95\% conf.int. \\ [0.5ex] 
 \hline\hline
Model 1     &      & Accuracy: 0.971& Pseudo R-squared:0.21  \\ \hline
Normalized Google difference & 2.44 & 0.0000 & [1.68, 3.20] \\ \hline
Google hits for male name   & 0.00  & 0.9949 & [0.00, 0.00] \\ \hline
(Intercept)   & 2.41  & 0.0000 & [1.92, 2.90]    
\\ [0.5ex] 
 \hline\hline
Model 2  &      & Accuracy: 0.995 & Pseudo R-squared:0.62 \\ \hline
Normalized Google difference & -5.93& 0.006 & [-10.13, -1.73] \\ \hline
Google hits for male name   & -2.07e-5  & 0.558  & [-9.02e-3, 4.87e-5]    \\ \hline
(Intercept)   & -5-55  & 0.001  & [-8.87, -2.24]     \\ \hline
  
\end{tabular}
}
\end{table}

%REPETITIOn???
%Our findings reveal that the German speaking web is biased towards male job titles, meaning that for most of the professions one can find much more sources for male than female job titles. 
%The results of analysis of professions with Wikipedia articles reveal that professions from the female bias group have a significantly lower normalized Google difference than the professions in the male bias group. In other words, Google hits are lower for male than corresponding female job titles in professions that have only an article with a female job title as an article title. At the same time, Google hits are greater for male than female job titles in professions that have only an article with a male title. However, the balanced bias group does not show the significant difference with the male and female bias groups. The normalized Google difference of the balanced group varies from 1 to -0.6 %(Figure 7)
%, which means that professions in the group can have popular either female or male job titles. 

\textbf{Labor Market.} Next we investigated if the gender over- or under-representation in profession titles on Wikipedia can be explained by labor market statistics. We hypothesized that professions which are dominated by women nowadays will be found in the ``female bias'' group, whereas professions dominated by men will be found in the the ``male bias'' group we created. 
To this end, the dependence between the percentage of women involved in a profession and the redirection bias was analyzed. 
%Figure 8 shows the percentage of women involved in professions; data is grouped by redirection bias. One can see that professions in the female bias group have between 82 and 100 percent of women (median is 93.7) and the male bias professions have from 0 to 100 percent of women, where half of professions in male bias group have between 15 and 60 percent of women. The neutral and male bias groups show similar distributions of women involved in a profession, with slightly different median values, 28.3 and 35.5 percent of women, respectively. This suggests that potentially there is no significant difference between groups of male and neutral bias in terms of employed women.
%
%Next the dependence was tested between the percentage of employed women in the German labor market and the redirection bias. The results of Wilcoxon rank-sum tests suggest a statistically significant difference (the alpha according to the correction method is 0.0167) between the underlying distributions of:
%1)	the percentage of women involved in male bias professions and the percentage of women involved in female bias professions (z = -3.32, p<0.001);
%2)	the percentage of women involved in neutral bias professions and the percentage of women involved in female bias professions (z = -3.35, p<0.001).
We observed statistically significant differences in terms of percentage of women involved in a profession between: (1) the male and female bias groups ($z = -3.32, p^{**}%<0.001
$), and (2) neutral and female bias groups ($z = -3.35, p^{**}%<0.001
$). In other words, professions which are represented by only articles with female titles (i.e., female bias) show a significantly higher percentage of employed women than other professions, i.e., between 82 and 100 percent of women, whereas professions that are represented by only articles with male title tend to have from 10 to 60 percent of women. 
There was no significant difference between the underlying distributions of the percentage of women involved in professions which are in the male bias and neutral groups.% (z = 0.83, p> 0.1). 

\begin{figure}[t!]
    \centering
    \includegraphics[width=0.8\linewidth, trim=280 12 60 25, clip=true]{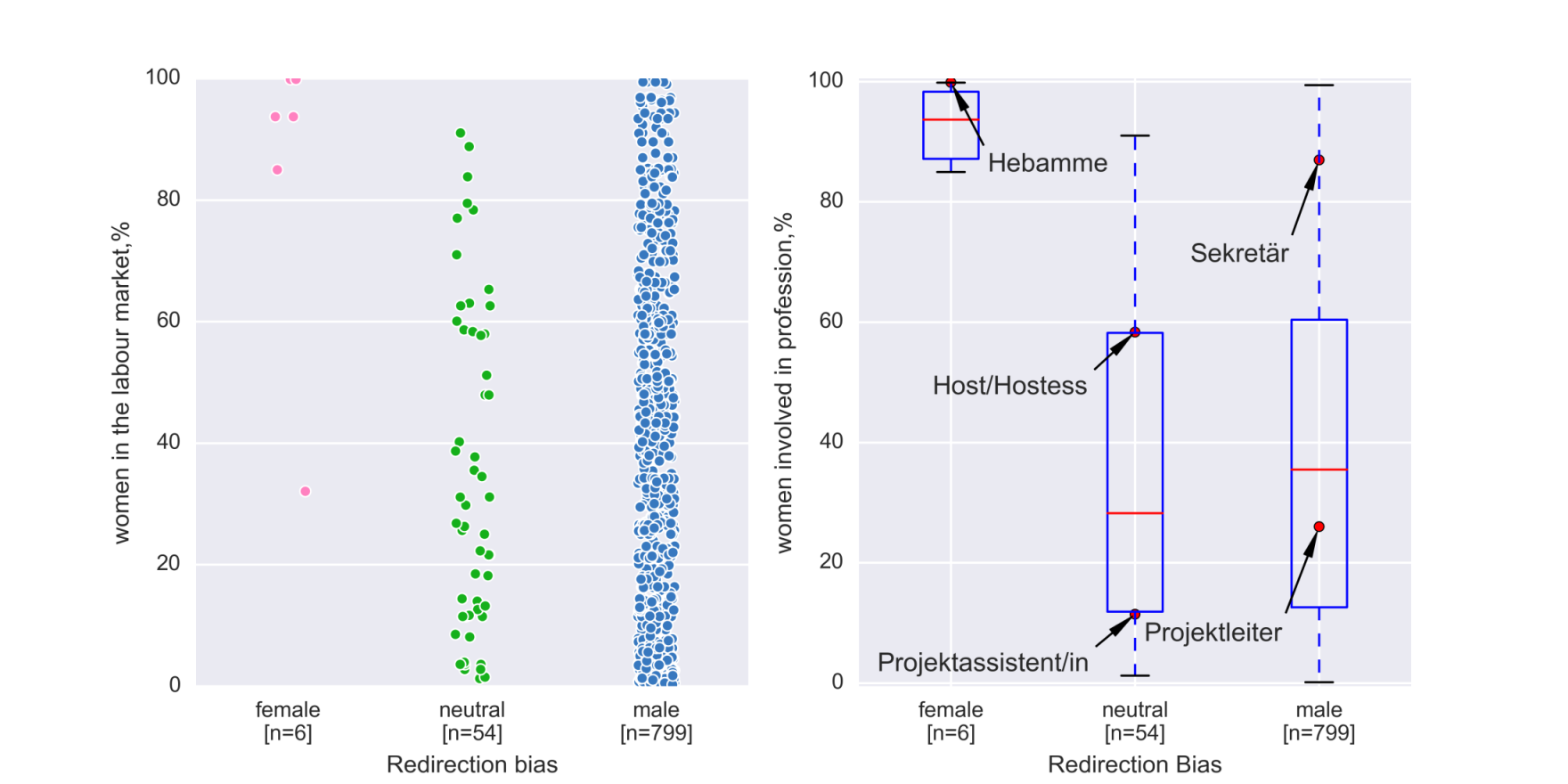}
    \caption[Percentage of women involved in professions]{\textbf{Percentage of women involved in professions} (German labor market).
 Data is grouped by redirection bias of professions. The statistics from the German labor market were associated with 859 professions (799  male  bias, 54 neutral,  and 6  female bias). For some generic labels (e.g., DE: ``Leiter'', EN: ``head'', ``leader''), no labor market statistics exist. 
One can see that professions with a female redirection bias on Wikipedia are indeed dominated by women in the German labor market, while there is no clear relation with the labor market statistics for the other two groups.}
    \label{fig:labor_bias}
    \vspace{-3mm}
\end{figure}

Nevertheless, one can glean from  Figure \ref{fig:labor_bias} that some professions with 80-100 percent of women are in the male bias group. For example, in the profession “Gesundheits- und KrankenpflegerIn” (EN: health and nursing staff) one encounters about 85\% women and the Wikipedia page with the female title “Gesundheits- und Krankenpflegerin” automatically redirects to the respective male form “Gesundheits- und Krankenpfleger”, hence the profession is in the male bias group. Another example is profession “Sekretär/in” (EN: secretary) whose workforce is constituted of about 88\% women in Germany and only features the male article “Sekretär”. Thus, we cannot claim that the Wikipedia community, e.g., decided to use only female profession names as article titles for professions with a majority of women in the labor market. We can only conclude that existing professions with only a female profession article are more likely to have higher percentage of women in the labor market than professions with only a male profession article.

Second, the logistic regression model was fitted in order to describe relations between the redirection bias groups using the percentage of women involved. Because professions from the neutral and male bias groups do not show significant differences, we fit a logistic regression which will predict whether a profession is in the female bias group or not -- i.e., the neutral and male bias groups were processed together. 
%
%\begin{figure}[t!]
%    \centering
%    \includegraphics[width=\linewidth, trim=0 0 0 0, clip=true]{figures/log_reg_labormarket.png}
%    \caption[Logistic regression Labor market]{
%Results of the best fitted logistic regression model with the binary dependent variable of a profession being in the female bias group. The coefficient for “Percentage of involved women” reveals that, we will see 44\% increase in the odds of being in a female bias group for a one-unit increase in percentage of involved women, since $exp(0.364) = 1.44$. For example, probability of a profession being in the female bias group for a profession with 20\% women in Germany equals $5.3e-13$ and $0.55$ if a profession has 98\% women.  
%    }
%    \label{fig:log_reg_labor}
%\end{figure}
%
\begin{table}[b]
\centering
    \caption[Logistic regression Labor market]{\textbf{Logistic regression for the labor market.}
(Results of the best fitted logistic regression model). The binary dependent variable represents a profession being in the female bias group. The coefficient for “Percentage of involved women” reveals that we will see a 44\% increase in the odds of being in a female bias group for a one-unit increase in percentage of involved women, since $exp(0.364) = 1.44$. For example, the probability of a profession being in the female bias group for a profession with 20\% women in Germany equals $5.3e-13$ and $0.55$ if a profession has 98\% women.  
    }
\label{log_reg_labor}
\resizebox{\columnwidth}{!}{
\begin{tabular}{|c|c|c|c|}
\hline
    & coef.  & p & 95\% conf.int. \\ [0.5ex] 
 \hline\hline
    &      & Accuracy: 0.99& Pseudo R-squared:0.61  \\ \hline
Percentage of involved women & 0.36 & 0.008 & [0.10, 0.63] \\ \hline
(Intercept)   & -35.53  & 0.005 & [-60.44, -10.62]    
\\ \hline
  
\end{tabular}
}
\end{table}
The logistic regression coefficients (Table \ref{log_reg_labor}) reveal that a one-unit increase in the percentage of employed women will result in a 44\% increase in the odds of being in the female bias group (versus being in the male bias or neutral groups). %In other words, increasing the percentage of employed women leads to a probability increase of a profession being in the female bias group.
The output indicates that the percentage of employed women is significantly associated with the probability of being in the female bias group, yet this relation shows itself only at very high values of female employment ratios. For example, we have an estimated probability of 0.001 for being in the female bias group for a profession with 80\% employed women. The estimated probability is instead 0.63 for a profession with 99\% employed women. That means that only professions with very high (> 97\%) percentages of employed women will be likely to have only an article with a female title. %Otherwise professions are more likely to have articles with a neutral or male title or they have articles with both male and female title.

Results of rank-sum tests and fitted logistic regression model support the hypothesis that professions with higher percentages of women have only a female page (i.e., professions from female bias redirection group). At the same time, the threshold of percentage of employed women, where one observes increase in likelihood of profession to be in the female bias, is very high (about 97\% employed women).
However, we cannot identify a significant difference between the male and neutral bias groups. The relation between these groups cannot be described using the percentage of employed women in the labor market. 

\subsection{Images analysis}

%From the corpus of profession articles we extracted 906 images which belong to 345 profession articles. These images were classified by workers of a designated CrowdFlower task. Each image was assessed by at least three workers.

%Workers identified whether each photo shows people or not. According to the results of the survey, 34.3\% of images do not show people on it. For every image with exactly one person (31 \%), workers were asked to identify the gender. If an image showed more than one person, then workers were asked to decide whether one person is dominant (more details in Subsection 4.2) so that they could identify the gender only for that dominant person (5.8\%). Otherwise, the workers were asked to identify the gender of the majority on the photo, or if the gender is not recognizable or the photo shows about the equal number of men and women, then they should choose corresponding variants. Results reveal that there were 27.3\% of images with several persons without a dominant person. 
The aggregated results for all 906 labeled images were %represented in Figure 9,
 as follows: 
%where “women in image” and “men in image” corresponds to one woman/man or majority of woman/man at photos with several persons without a dominant person. One can see from Figure 9 that 
(1) almost half (44.8\%) of the images show men, whereas only 12.4\% of images show women;
(2) an equal amount of men and women is observed only in a few images (5.1\%);
(3) almost a third (34.3\%) of the images depicts no person; 
(4) gender is not recognizable for 3.4\% of the images.

\begin{figure}[t!]
    \centering
    \includegraphics[width=\linewidth, trim=0 0 0 0, clip=true]{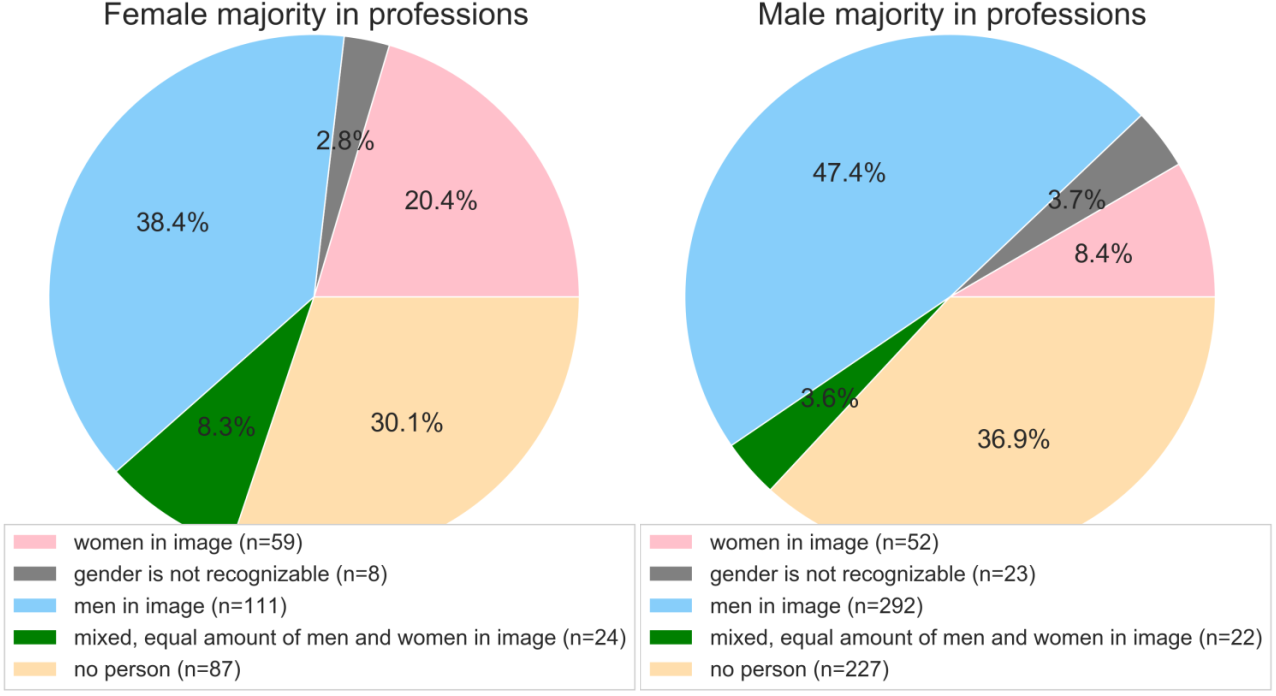}
    \caption[Image categories, grouped by gender of majority]{
\textbf{Distribution of image categories, grouped by gender of majority in the labor market}. Professions with a female employment majority have almost two times more images depicting men than women, whereas professions with male majority feature even fewer images depicting women, around 5.6 times less.
    }
    \label{fig:images_labor50}
    \vspace{-4mm}
\end{figure}

The analysis of images, grouped by gender of article title (profession name) where the images were encountered reveals a significant difference in image composition. Articles with a female title have almost 10 times more images depicting women than men, whereas articles with a male title have 4 times more images depicting men than women. Articles with a neutral title have 6 times more images depicting men than women. Thus, articles with female and male titles both show a respective gender inclination. While men are vastly underrepresented in articles with female titles, the under-representation of women in articles with male and neutral titles exists, but is much weaker. %, meaning that almost 4 times more images depicting men than women were observed. 
We observed the same tendencies when grouping images by redirection bias of profession. Professions with only an article with a female title on Wikipedia have most images depicting women (75\% of images). Nevertheless, professions from “male bias” and “neutral bias” redirection groups are not gender neutral in terms of image representations, as the majority of images depicts men (44.4\% and 52\% of all images, correspondingly).

%It would be interesting to know how composition of images relates to labor market statistics. Thus, the following subsection represents a comparison of distributions of image categories with respect to labor market data as well as an analysis of the relation between the number of images depicting men/women and the labor market statistics.

\textbf{Labor Market.} %To test whether images from the profession articles reflect labor market statistics, an analysis of image groups was performed. 
Professions were divided into two groups: professions with female majority and such with male majority (>50\%). %Thus, professions with more than 50\% women were in one group and professions with more men were in another group. 
%
%Naturally, if profession images on Wikipedia reflect the German labor market statistics, one will observe the larger part of images depicting women in the group of professions with female majority and the larger part of images depicting men in the group of professions with male majority. 
%
%
Figure \ref{fig:images_labor50} shows the distribution of image categories, grouped by gender of majority in the labor market. One can see that there are almost two times more images depicting men (38.4\%) than women (20.4\%) in professions with more than 50\% women in labor market (female majority group). At the same time, there are almost 6 times more images depicting men (47.4\%) than women (8.4\%) in professions with male majority. 
In both groups we observe that the majority of images depict men. Therefore, we can conclude that profession images on Wikipedia do not reflect the labor market statistics. 
We also looked at groups with more than 70\% men and women in the labor market.%\footnote{Other groups can be found at \url{https://github.com/gesiscss/Wikipedia-Language-Olga-master/blob/master/9.3\%20Analysis\%20of\%20images.ipynb }}.

%Next, professions were divided in two groups: professions with at least 70\% men and at least 70\% women in the labor market. Figure 13 represents the distribution of image categories among these groups. One can see that there is almost an equal number of images depicting men and women (33.0\% and 32.1\%) in the group of professions with at least 70\% women in the labor market. In contrast, professions with at least 70\% men have about 7 times more images depicting men than women (48.7\% vs 6.7\%). 

%We found a statistically significant relationship (p<<0.0001)) between the category of images and the dominant gender of the profession. In other words, a significant difference in the image categories composition was observed between the female- and male-dominated professions. According to the multiple post-hoc tests, images depicting women (p<<0.0001), images depicting men (p<0.01), images with equal number of men and women (p<0.01), and images without people (p<0.01) showed a significant difference between the female- and male-dominated professions.

An analysis of the images, grouped by gender of majority (>50\%) %and by dominating gender (>70\%),
according to the labor market statistics, reveals a significant difference in image composition. Professions with more than 50\% women in the labor market have 2 times less images depicting women than men, whereas professions with more than 50\% men in the labor market have almost 6 times less images depicting women than men. In other words, a majority of one gender in the labor market does not imply a majority of the same gender in images. %The image majority remains biased against women. If professions are limited to those which have in the labor market at least 70\% persons of one gender, an equal number of images will be encountered in articles of female-dominated professions and gender inequality (bias against women) in articles of male-dominated professions.

\begin{table}[t]
\centering
\caption{Spearman’s rank correlation coefficients between the number of images in Wikipedia article about a profession and the labor market statistics of the profession. }
\label{table-labor-spearman}
\resizebox{\columnwidth}{!}{
\begin{tabular}{|l|l|c|}
\hline
   Images    & Labor market  & Corr. \\ [0.5ex] 
 \hline\hline
%number of img depicting women     & number of women     & 0.15*       \\ \hline
%number of img depicting men       & number of men       & 0.088       \\ \hline
%percentage of img depicting men   & percentage of men  & 0.3***      \\ \hline
percentage of img depicting men   & percentage of women  & -0.3***     \\ \hline
percentage of img depicting women & percentage of women & 0.34***     \\ \hline
%percentage of img depicting women & percentage of men   & -0.34***    \\ \hline
%number of img depicting women     & percentage of women & 0.34***     \\ \hline
%number of img depicting women     & number of people    & -0.002      \\ \hline
%number of img depicting women     & number of men      & -0.1.       \\ \hline
%number of img depicting men       & percentage of men    & 0.2**       \\ \hline
%number of img depicting men       & number of people     & 0.03        \\ \hline
%number of img depicting men       & number of women    & -0.06       \\ \hline
%percentage of img depicting men   & number of people   & -0.07       \\ \hline
%percentage of img depicting women & number of people   & 0.01        \\ \hline
percentage of img depicting women & number of women    & 0.17**      \\ \hline
%percentage of img depicting women & number of men     & -0.09       \\ \hline  
percentage of img depicting men & number of men     & 0.013       \\ \hline
\end{tabular}
}
\end{table}

%Finally, we want to know if the number of images depicting men/women relates to the labor market statistics. So, numbers of images in profession articles were correlated with the German labor market statistics for professions. Table \ref{table-labor-spearman} represents Spearman’s rank correlation coefficients. 

Results of Spearman’s rank correlation (Table \ref{table-labor-spearman}) reveal a moderate positive correlation ($\rho=0.34$) between the percentage of images depicting women in an article and the percentage of women in the corresponding profession, and a moderate positive correlation ($\rho=0.3$) between the percentage of images depicting men in an article and the percentage of men in the profession. 
Thus, professions with higher percentage of images depicting women/men
exhibit a moderately higher percentage of women/men in the labor market. %At the same time, there is no correlation between the percentage of depicted men and the total number of employed men in a profession.

\subsection{Analysis of mentioned names}

There are 411 articles about professions which mention at least one person. The articles mention overall 5085 persons (4272 men and 813 women). 10.4 men and 1.9 women were mentioned on average in the articles. %For each article the proportion of mentioned men was estimated; 
The mean proportion of mentioned men per article is 0.83 and the median proportion is 0.98. %(Figure \ref{fig:people_all}). 
In other words, out of all people mentioned in an article, 83\% of them are men and only 17\% of them are women on average.

\begin{figure}[t!]
    \centering
    \includegraphics[width=0.8\linewidth, trim=7 8 0 15, clip=true]{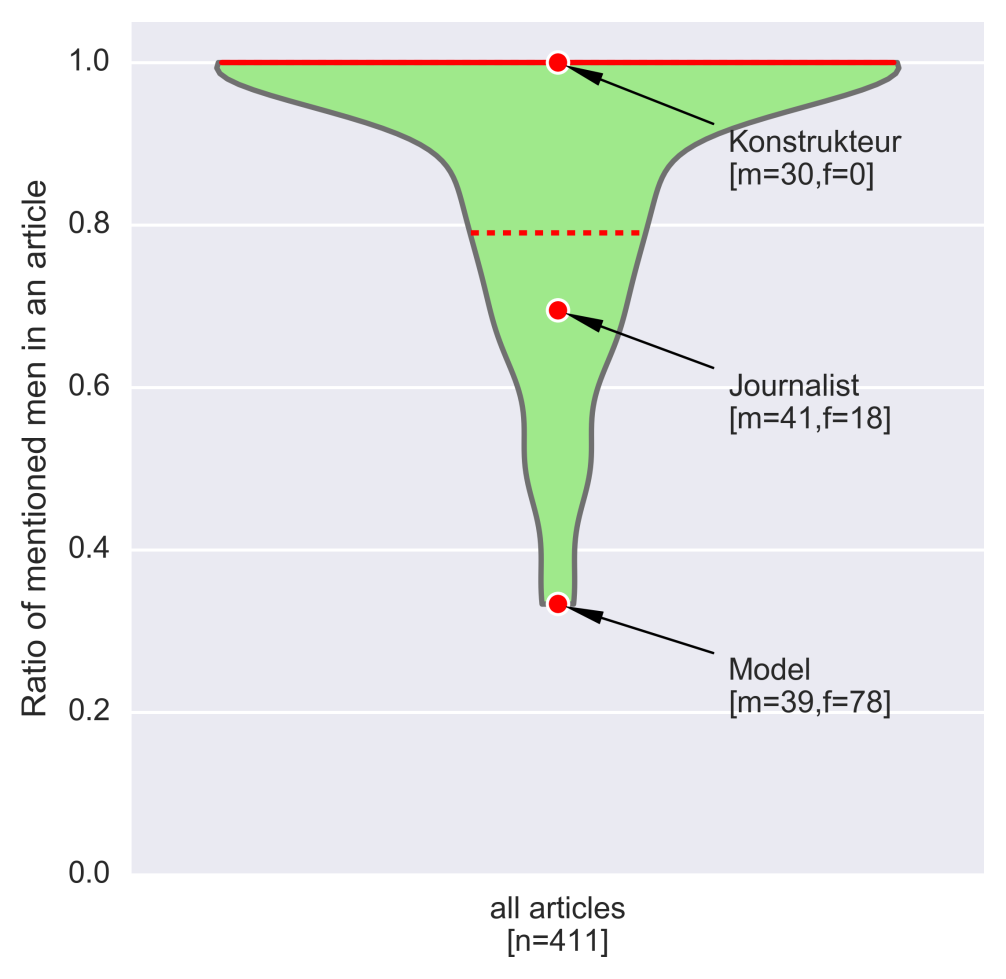}
    \caption[Mentioned men proportions]{\textbf{Mentioned men proportions.}
Violin plot of the distribution of proportional mentions of male names in articles. The violin plot features a kernel density estimation of the underlying distribution. The shape of the violin plot represents the density: the wider the violin plot, the greater the number of articles with a particular proportion of men. The solid red line represents the median; the dotted line shows the 3rd quartile (1st quartile line hidden). The violin plot represents data points which are distributed within two standard deviations of the mean, outliers are removed. One can see that ½ of articles mention only men in the articles, ¾ of articles mention 80-100\% men and only 0-20\% women. Moreover, only a few articles mention either an equal number of men and women or more women than men. %For example, the article “Model” mentions 39 men and 78 women.
    }
    \label{fig:people_all}
    \vspace{-3mm}
\end{figure}

Figure \ref{fig:people_all} shows the distribution of men proportions. As one can see, 50\% of articles have a proportion of mentioned men close to 1.0 and 75\% of articles have a proportion of mentioned men from 0.8 to 1.0. %Moreover, Figure \ref{fig:people_all} features the following examples: 
The article “Konstrukteur” (EN$\approx$ ``design engineer'') for example mentions 30 persons, all of them men, the article “Journalist” mentions 41 men and 18 women, and the article “Model” 39 men and 78 women.
%
%\wac{I would remove this par. too much detail again}
%Since articles with only one (or two) mentioned men will have proportion of mentioned men equal 1.0, one might assume that proportion of mentioned men is a misleading measure. In order to control that effect, the relations were examined between the proportion of mentioned men and the following: i) number of mentioned men, ii) number of mentioned women. (Alternatively, we could also remove from an analysis all articles with very few mentioned persons.) A positive weak correlation (Pearson cor.coef. 0.11) was found between the proportion of mentioned men and the number of mentioned men, and a negative correlation (Pearson cor.coef. - 0.29) was found between the proportion of mentioned men and the number of mentioned women. In other words, articles that mention more men have a slightly higher proportion of mentioned men and articles with higher proportion of men mention a slightly lower number of women. 
%
Articles which mention an equal number of men and women %(i.e., the proportion of mentioned men equals $0.5\pm0.05$) 
are considered gender neutral. Consequently, all articles with $>50\%$ male names are male-biased with the remainder being female-biased in our consideration.
%proportions of mentioned men higher than $0.5+0.05$ are male biased and all articles with the proportion smaller than $0.5-0.05$ are female biased. 
One can see from Figure \ref{fig:people_all} that most articles have a proportion higher than 0.5. %In other words, male bias is observed in most articles. 
In particular, 92.5\% of the articles are male-biased, 3.1\% of the articles are female-biased, and 4.4\% of the articles are gender-equal.

Lastly, we grouped articles according to the gender of the article title and found that the distribution of proportions of mentioned men was significantly different ($p^{*}%p<0.05
$) between articles with male and female titles.   
%Next, the dependency was tested between the proportion of mentioned men and the gender of the article title where these men were mentioned. Basically, we want to know if: i) articles with a male title have a higher proportion of mentioned men than articles with a neutral or a female title; ii) articles with a neutral title have a higher proportion of mentioned men than articles with a female title. 
%
%First, the articles were grouped according to gender of article title (Figure 15). Then three rank-sum tests were performed in order to compare proportions of mentioned men of the groups. Results of the tests reveal that there is a significant difference between distribution of values in the articles with male versus female title ($p^{*}%p<0.05
%$). 
Proportions of mentioned men in articles with male titles (median is 0.98) are more likely to be larger than the proportions in articles with female title (median is 0.65).% There is no significant difference between pairs of groups: male-neutral, female-neutral.

\textbf{Labor Market.}
%Next the relation was examined between the labor market statistics of a profession and the number of mentioned men/women in the profession article on Wikipedia. To begin, t
The difference in proportions of mentioned men was explored between professions with male and female majorities. % Then we tested the relations between the number of mentioned men/women and the number of employed men/women in a profession. 

%
%TODO: modify text. remove mentions of total numbers
%
%In order to test differences in proportions of mentioned men, the professions were divided in two groups: i) professions with more than 50\% employed women; ii) professions with more than 50\% employed men. %Figure 16 shows the distributions of mentioned men proportions in articles with male and female majority in the German labor market. Then the rank-sum test was performed between values of these two groups. 
%
%The analysis reveals a significant difference between the distributions of men proportions in professions with male and female majority ($p^{***}%p<<0.0001
%$). The proportions of mentioned men in the articles with male majority (median proportion is 0.98) are more likely to be larger than the proportions of mentioned men in the articles with female majority (median proportion is 0.88). In other words, professions with female majority have articles with a significantly lower proportion of mentioned men than articles of professions with male majority. 
%
%Nevertheless, 
Three quarters of articles
%(Figure 16)
among professions with female majority in the labor market mention 68\%-100\% men and 0-32\% women. This implies that the largest part of the articles for professions with female majority expose male bias; this is despite the fact that compared to professions with male majority, female majority articles still feature a significantly lower proportion of mentioned men (medians are 0.88 and 0.98, $p^{***}$ for male majority professions) .

%Next we want to know the strength of the relation between the percentage of mentioned men/women in the article and the percentage of employed men/women in the profession. 
Results of Spearman’s rank correlation (Table \ref{table-labor-mentions-spearman}) reveal weak positive correlation ($\rho=0.27$) between the percentage of mentioned women or men and the percentage of women or men in the labor market correspondingly. %The same correlation can be found for the percentage of mentioned men and the percentage of men in the labor market. 
In other words, the higher is the percentage of women in a profession, the higher is the percentage of mentioned women in the article about the profession, and the other way around. In summary, one can conclude that for some professions the percentage of mentioned men or women reflects the labor market statistics in terms of gender proportions.

\begin{table}[t]
\centering
\caption{Spearman’s rank correlation coefficients between the number of mentioned men/women in a Wikipedia article about a profession and the labor market statistics of the profession. }
\label{table-labor-mentions-spearman}
\resizebox{\columnwidth}{!}{
\begin{tabular}{|l|l|c|}
\hline
   Mentioned people    & Labor market  & Corr. \\ [0.5ex] 
 \hline\hline
percentage of mentioned women & percentage of women  & 0.27***     \\ \hline
percentage of mentioned men   & percentage of men   & 0.27***     \\ \hline
%number of mentioned men       & number of men       & -0.23***    \\ \hline
%number of mentioned men       & number of women     & -0.15**     \\ \hline
%number of mentioned men       & number of people    & -0.20***    \\ \hline
%number of mentioned women     & number of men     & -0.08       \\ \hline
%number of mentioned women     & number of women    & 0.09.       \\ \hline
%number of mentioned women     & number of people   & -0.01       \\ \hline
%number of mentioned persons   & number of men       & -0.21***    \\ \hline
%number of mentioned persons   & number of women     & -0.09.      \\ \hline
%number of mentioned persons   & number of people   & -0.17***    \\ \hline
%number of mentioned women     & percentage of women  & 0.26***     \\ \hline
%number of mentioned men       & percentage of men   & -0.10*      \\ \hline
percentage of mentioned women & number of women     & 0.17***     \\ \hline
\end{tabular}
}
\end{table}

\section{Related work}\label{sec:relatedwork}

%\textbf {Gender bias in Wikipedia and on the Web.}
Due to the increasing importance of online media, much research is concerned with an assessment of bias in Wikipedia and on the Web in general. %A lot of attention was drawn towards biases in Wikipedia and especially gender bias.
For example, Reagle and Rhue \cite{reagle_gender_2011} compared biographical articles from the English Wikipedia edition and the online Encyclopedia Britannica with respect to coverage, gender representation, and article length. Authors concluded that Wikipedia provided better coverage and longer articles. While Wikipedia has more articles on women than Britannica in absolute terms, Wikipedia articles on women are missing more often than are articles on men, when compared to Britannica. Wagner et al.\cite {wagner_its_2015} studied coverage of famous women in Wikipedia articles and the way women are portrayed in the online encyclopedia. The authors found that, despite good coverage of famous women in many Wikipedia language editions, the ways in which women and men are portrayed differ significantly. For example, romantic relationships and family-related issues are much more frequently discussed about women than men. %Graells-Garrido et al.
In
\cite{graells-garrido_first_2015,wagner_its_2015,wagner_women_2016} differences between descriptions of male and female biographies in terms of network structure, 
topical focus, structural properties
and language were researched. %Researchers revealed that articles about men are disproportionately more central than articles about women and the words most associated with men are about sports, while the words most associated with women are about arts, gender, and family.
%Further research \cite{wagner_women_2016} was extended to notability, topical focus, linguistic bias, structural properties, and meta-data presentation which systematize and modify approaches used in \cite{wagner_its_2015,graells-garrido_first_2015}.
Thus, several studies suggest that gender bias on Wikipedia can be assessed using articles with biographies. %However, neither of these approaches and measurements can reveal gender bias related to professions.
%Another branch of research focuses on the gender gap among Wikipedia editors. %Two groups of studies exist: (1) studies focused on reasons, and consequences of gender bias;
%(2) studies focused on measuring the gender gap. 

Several studies \cite{collier_conflict_2012,lam_wp:clubhouse:_2011} have tried to explain gender imbalance by studying how conflict-related behaviors (e.g., reverts) affect male and female editors. % in order to understand why an imbalance might exist
%For example, authors found that female editors are reverted more than males. 
Collier and Bear \cite{collier_conflict_2012} studied reasons for why female contributors stop contributing, indicating that the gender contribution gap is due to responses to conflicts.
In \cite{hargittai_mind_2015, hinnosaar_gender_2015} authors found that significant Internet experiences and beliefs about one’s competence explain a large share of the gender gap. 
%and skills help explain a critical aspect of the variation in the contribution rates among men and women. %Thus, higher levels of Internet skills predict much greater probability of contribution for men than women. However, low-skilled men and low-skilled women are equally high unlikely to contribute to Wikipedia.
%Hinosaar\cite{hinnosaar_gender_2015} manifested that gender differences in the frequency of Wikipedia use and beliefs about one’s competence explain a large share of the gender gap. 
 According to \cite{hill_wikipedia_2013} %,_wikipedia_2011}%, only 9\% of contributors are women. In particular, they encountered from 3\% to 20\% women in top 10 countries of editors’ residence. Hill and Shaw\cite{hill_wikipedia_2013} modified the technique used for the estimation of female proportion. While researchers reported slightly higher proportion, i.e., 
only 16.1\% of editors on Wikipedia are women.
Antin et al.\cite{antin_gender_2011} studied differences between men and women’s editing activity in terms of the number and size of the revisions they make. %They found that among the most active Wikipedians men tended to make more revisions than women. However, they also found that the most active women in the sample tended to make larger revisions than the most active men.
Bernacchi \cite{viola_bernacchi_gender_2015} performed a visual exploration of the gender issue on Wikipedia by studying the articles “Man” and “Woman” in several language editions of Wikipedia. The researchers compared sizes of articles, numbers of edits per article, changes in TOC structures, network structures of related articles and intersection of used concepts, topics coverage, revisions in terms of vandalism and deleted content.
However, profession-related gender bias on Wikipedia and corresponding aspects  %(e.g., images and mentioned people) 
remain uncovered, to the best of our knowledge.

%\textbf {Gender bias in image search results and traditional media.}
Kay et al.~\cite{kay_unequal_2015} studied the gender bias in image search results % for several professions. They 
and found that manipulated search results have a small significant effect on people’s gender ratio perception  (cf. Section \ref{sec:intro}). 

\section{Discussion}\label{sec:discusion}

In general, we observe a strong overrepresentation of male titles, images and mentioned names on the German Wikipedia that cannot be explained in its entirety by labor market statistics. 

%For most professions one can only find an article in the German Wikipedia that has the male job name as the article title. 
The analysis of article titles and redirections reveals that most professions are represented only via an article with a male form of the profession. Moreover, most encountered redirections are from female to male titles. This evidence supports the existence of gender disparity along article titles in the professional domain. 
Choices of article titles mostly reflect the general popularity of male over female profession names on the German speaking Web, with some exceptions. %In other words, Wikipedia reflects the general gender bias observed on the Web. %Professions represented only by an article with female title, are more likely to have more sources for female than male job titles according to the number of Google results for the title. Professions, which are represented only by articles with male titles, are more likely to have more sources for male job titles. 
Turning to the labor market statistics, we observe a relation between the percentage of women involved in a profession and the probability of having only a female article on Wikipedia. However, only professions with a distinctly larger female ratio in the workforce are likely to be represented only by a female article. On the other hand, using labor market statistics, we cannot well distinguish between professions with both male and female lemmata versus professions represented only by a male title, as the male redirection/representation bias on Wikipedia does not only appear for ``male'' professions as measured by true employment. 

Regarding depictions, almost four times more images of men than women are used to describe the professions. We find that almost half of the images from the profession articles depict men and only around 12\% show women. %Thus, we encounter male bias in profession images on Wikipedia. 
We observe a significantly higher number of images depicting men in professions, regardless of female majority or male majority in the labor market for that profession. 
%Hence, we encounter evidence for male bias within these profession groups.
Only for articles under female lemmas and professions which have only a female article on Wikipedia do we find a significantly higher number of images depicting women than men. %Hence, we encounter evidence for female bias within these professions with respect to images depicting women. 

Along the mentioned people dimension, an imbalance towards male names (more than 50\% men) was observed in 92.5\% of the articles. %Generally, more men than women were mentioned in most articles.
Only very few articles mention exhibit either (a near) equal number of men and women or more women than men. Particularly, out of all mentioned people, we observed ¾ of articles with 79-100\% men and 0-21\% women. Less than 50\% men (i.e., female bias) were mentioned in 3.1\% of the articles and equal representation was observed in 4.4\%. %If we restricted mentioned people to those who were born after 1960, we observed 50-100\% men in ¾ of articles and 100\% men in ½ of articles. Considering the percentage of articles with (near to) equal number of genders, a male bias was observed in approx. 72\% of articles, gender equality was observed in almost 7\% of articles and a female bias was observed in approx. 22\% of articles. Thus, we can conclude that elimination of people, who were born before 1960, leads to a less skewed gender inequality towards male bias, meaning that the percentage of articles with male bias decreases. However, male bias remains in approx. 72\% of articles. 
Analyses of the articles with female title and articles of professions with female majority in the labor market revealed significantly lower proportions of mentioned men. Nevertheless, even for groups where the difference of proportions was significant, a male bias was observed in the larger part of articles from these profession groups.

Interpreting these results, one major reason for why male lemmas and redirects dominate in (the profession articles of) the German Wikipedia is certainly the existence of the ``Generisches Maskulinum'', i.e., the traditional German language tendency to refer to a group of mixed or unknown gender with the male descriptor. We have also found several discussions on the talk pages of the respective articles studied here that point to a deliberate decision to ``simplify'' the navigation structure of Wikipedia by agreeing on the guideline to put all profession names under their generic masculine name. This ``bias'' as we have called it hence doesn't automatically indicate a deliberate prejudiced ideology against female descriptors. Yet, the effect on a reader is the same: she is being redirected or cannot find results for female profession name forms, which certainly influences her perception of  how society views certain professions; also reinforcing the use of the generic masculine form that has been criticized in the German-speaking public discourse. A simple solution would be to host all articles under a gender-neutral lemma (as is practice already for some profession as we have seen), reachable via redirect from the male and female lemmas - the description of the profession througout the article then has to be simply adapted to a neutral/both-gender form (e.g., with the ``/In'' suffix). %We could not discern profound practical or technical reasons against implementing this solution from the talk page discussions.

Regarding images and names, another reason for male overrepresentation is presumably  that, especially in long-established professions like ``smith'', many older depictions  and records exist that contain for the most part men -- and as prominence often is achieved post-mortem, names of famous representatives of a trade are more likely to be mentioned by name in reference works, reflecting a historic imbalance towards working men. Nonetheless, the biases we unveiled also appear in some more modern professions. And, if we again consider solely the effect on the reader, in such cases a more recent focus on the occupational field might be useful, for example by concentrating the lede and starting sections on the contemporary state of a profession before covering its history. %Future research could shed light on such possible historical or "grammatical" biases. %Due to the fact that, in the past, women had a secondary role in science, culture, and history, we may observe much less women in professional domains.
To control for history effects, future work might filter out entity-matched persons by their date of birth, and specifically identify paragraphs that refer to historical aspects of a profession. %ord perform additional analysis for the extracted text and the remaining text. 

Regarding possible reasons for the observed imbalances, previous research~\cite{collier_conflict_2012,hill_wikipedia_2013} has revealed that a dominant majority of editors in Wikipedia is male. % Hill and Shaw \cite{hill_wikipedia_2013} reported 22.7\% US female editors and 16.1\% overall female editors on Wikipedia, which provides evidence for gender inequality in the Wikipedia editors community.
Wikipedia further contains fewer and less extensive articles about women or topics important to women~\cite{lam_wp:clubhouse:_2011}, indicating that male editors might not be suited to entirely represent the female world view and interests. Over the last  years, the Wikimedia Foundation has made many attempts to attract more female editors \cite{ciampaglia_moodbar:_2015,morgan_tea_2013}, although with mixed success.
%Another reason, which potentially explains the male bias on Wikipedia, is the male bias in other media (e.g., the Web). 
Yet, the gender of editors is likely not the only factor behind the described imbalances. Just as many historical reference works already contain a male-dominant description, Wikipedia might reflect and in some cases inherit biases from other media and the references contained therein. We observed evidence for a male bias on the German speaking Web, i.e., for most of professions one can find much more sources for male than female professions. Moreover, several studies \cite{fu_tie-breaker:_2016,%higgs_gender_2003,
smith_gender_2010} reveal gender bias and stereotypes in mass media. % and interviews of sport competitions.% Kay et al.\cite{kay_unequal_2015}. 
%Even though the Wikimedia Foundation was not reporting about success or failure of their attempts, attraction of more female editors is still needed.
%
%On the other hand, not only female editors can change the article content and make it more gender neutral.
I.e., apart from gender equality of the editor base, having transparent guidelines and rules towards representative equality in article content might prove just as useful to decrease gender disparities of individual articles; defining clear target audience groups and contemplating suitability of the content for those readers might likewise be helpful.

The method proposed in our research can in principle be applied for the analysis of gender inequalities in different Wikipedia editions. The image and mentioned people analysis methods can be applied to any language; the image analysis could in future work even be fully automated given the availability of reliable algorithmic tools. And while the redirection analysis can only be applied to languages with masculine and feminine grammatical genders, this leaves several candidate languages. A cross-language analysis extension is a promising goal for future work, by matching the respective profession pages, in order to compare biases over editions. Results could be compared to external country-wise data like the Global Gender Gap Index %\cite{schwab_global_2015}
for each country associated with a Wikipedia edition.

\section{Conclusion}\label{sec:conclusion}

%In this paper we presented an approach  

%We analyzed them over the following distinct dimensions: 
This study presents a  computational approach  to gender bias assessment along three distinct dimensions using Wikipedia profession articles. %The proposed approach may be applicable to other Wikipedia language editions. 
We made use of crowdsourcing to complement our analysis with high-quality data and compared against publicly available labor market data.  

The results indicate clear gender presentation imbalances on all the three dimensions studied (titles, mentioned persons, images). A notable portion cannot be explained simply by underlying labor market conditions, and many choices of representation seem to be made out of tradition or based on historic reference material. 

%We proposed a viable method for the identification and assessment of gender bias related to professions in Wikipedia articles. 
%The presented method can be implemented and incorporated into a software tool helping Wikipedia editors to identify and warn about existing gender inequality in articles along three dimensions: (i) gender-inclusiveness in job titles and corresponding redirection analysis, (ii) analysis of male-female image proportions, and (iii) the balance of male-female mentions.

This work has the potential to provide aides and inform guidelines for the Wikipedia community to identify and address gender disparities. Likewise, the outlined approach could be implemented in a software tool supporting Wikipedia editors in writing articles. The analysis procedure is likely reproducible for other Wikipedia language editions. Lastly, a careful consideration and possible evaluation of the effects of gender imbalance in profession articles especially on younger readers seems worthwile, given our results. 

The data and code are available online on GitHub.\footnote{\url{https://github.com/gesiscss/Wikipedia-Language-Olga-master/}}

\bibliographystyle{ACM-Reference-Format}
{\small
\bibliography{main}
}

\end{document}